\documentclass[12pt]{article}

\usepackage[english]{babel}
\usepackage[utf8x]{inputenc}
\usepackage[T1]{fontenc}

\usepackage{amssymb}
\usepackage{amsmath}
\usepackage{mathrsfs}
\usepackage{graphicx}
\usepackage{latexsym}
\usepackage{array}
\usepackage{multirow}
\usepackage[colorinlistoftodos]{todonotes}
\usepackage[colorlinks=true, allcolors=blue]{hyperref}
\usepackage{appendix}
\usepackage{verbatim}

\newtheorem{lemma}{Lemma}

\newcommand{\hsp}{\hspace{0.1in}}
\newcommand{\eqdef}{\stackrel{\triangle}{=}}

\newcommand{\GF}{\text{GF}}
\newcommand{\BB}{{\mathcal B} }

\begin{document}

\renewcommand{\baselinestretch}{1.2}

\title{\bf Scalable Block-Wise Product BCH Codes}
\author{Yingquan Wu and Eyal En Gad\thanks{The work was carried out when both authors worked at Micron Technology, Milpitas, CA 95035, USA} \\
}

\maketitle

\begin{abstract}
In this paper we comprehensively investigate block-wise product (BWP) BCH codes, wherein raw data
is arranged in the form of block-wise matrix and each row and column BCH codes
intersect on one data block. We first devise efficient BCH decoding algorithms,
including reduced-1-bit decoding, extra-1-bit list decoding, and extra-2-bit list
decoding. We next present a systematic construction of BWP-BCH codes upon given
message and parity lengths that takes into account for performance, implementation
and scalability, rather than focusing on a regularly defined BWP-BCH code. It can
easily accommodate different message length or parity length at minimal changes. It
employs extended BCH codes instead of BCH codes to reduce miscorrection rate and an
inner RS code to lower error floor. We also describe a high-speed scalable encoder.
We finally present a novel iterative decoding algorithm which is divided into three
phases. The first phase iteratively applies reduced BCH correction capabilities to
correct lightly corrupted rows/columns while suppressing miscorrection, until the
process stalls. The second phase iteratively decodes up to the designed correction
capabilities, until the process stalls. The last phase iteratively applies the
proposed list decoding in a novel manner which effectively determines the correct
candidate. The key idea is to use cross decoding upon each list candidate to pick
the candidate which enables the maximum number of successful cross decoding. Our
simulations show that the proposed algorithm provides a significant performance
boost compared to the state-of-the-art algorithms.

\end{abstract}

\section{Introduction}
\label{sec:intro}

Block-wise product (BWP) codes, particularly BWP-BCH codes, recently received quite a lot of research as well as practical interests \cite{Cho14} -- \cite{Kim16}.
In BWP codes, the user data is arranged in a two-dimensional array, composed of rows and columns. Each entry of the array is composed of multiple bits, and is called a {\em block}, or {\em intersecting block}, since it intersects a row and a column. Each row and column is encoded by an error-correcting code, typically a binary BCH code. In this work we only consider binary BCH codes as constituent codes. BWP-BCH codes are also called \emph{block-wise concatenated BCH (BC-BCH) codes}. The decoding of a BWP-BCH code is performed iteratively, such that in each iteration the constituent sensewords in one dimension (say, the rows) are decoded first, and subsequently the words of the opposite dimension (columns) are decoded. The expansion of the intersection from a single bit in conventional product codes to block in BWP-BCH codes allows for stronger (and fewer) constituent codes for the same overall block length and rate parameters. It turns out that having stronger constituent codes, even though their number is smaller, provides significant performance boost. A BWP-BCH code with this structure is said to be \emph{parallel concatenated}, since the rows and columns can be encoded in parallel. BCH codes can also be block-wise concatenated in series, by encoding one dimension first (say, the rows), and then encoding the opposite dimension, \emph{including} the parity bits of the first dimension. This is advantageous for classical product codes, but is less effective for BPC codes wherein the parity-on-parity property no longer holds. An optimization of the construction to improve the BCH parameters was proposed in~\cite{Kim15}. Serial concatenation is investigated in~\cite{Kim16}. The decoding of BWP-BCH codes can be performed in a soft or hard manner. In hard decoding, the decoder receives a binary channel output for each transmitted bit. The hard decoding process is simply to iteratively alternate row-wise and column-wise hard decoding using the Berlekamp algorithm. In soft decoding, the decoder receives multi-bit soft information regarding the likelihood of the channel value of each transmitted bit, wherein soft decoding of BCH codes is often referred to as Chase-II decoding \cite{Pyndiah98}.

An important challenge in the design of BWP-BCH codes and their decoder is the mitigation of error floor. An error floor forms in BWP-BCH codes when the noise level is low enough for most constituent words (rows and columns) to be corrected, except for a small number of words whose error count exceeds their correction capability.
Three methods were proposed to mitigate the error floor of BWP-BCH codes: soft decoding~\cite{Cho14}, a concatenation of an erasure code over the intersecting blocks~\cite{Yu14}, and collaborative decoding~\cite{Kim15}. In \cite{Cho14}, the decoder obtains additional soft information from the media and uses it to decode the failed BCH codes. \cite{Yu14} considers to concatenate an erasure code over the intersecting blocks, such that each block is treated as a symbol of the erasure code. Erroneous blocks are identified by the intersections of failed rows and columns, and are corrected by the outer erasure code. It explored Raptor or Reed-Solomon (RS) codes as erasure codes. In \cite{Kim15}, the row and column that intersect in a single erroneous block is combined to cancel the errors in the intersected block (but adding up errors in parity) in attempt to correct the errors.

BWP-BCH coding is a strong contender in solid-state-drive (SSD) controller \cite{Cho14, Yu14, Kim16}. It particularly appeals to enterprise SSD storage, wherein latency is the most critical metric. Normal read in flash NAND outputs single bit information without soft reliability, while it takes much longer latency to generate soft reliability (Typically multiple retry reads are carried to externally combine into multi-bit soft information, or a dedicated NAND command carries multiple reads and combines into multi-bit soft information internally). Thus soft-read in flash does not appeal to enterprise storage.  Data sector length is prevalently 4K bytes (possibly with extra metadata) while gradually migrating to 8K bytes. The typical parity overhead ranges 8 $\sim$ 16\%. The error floor is required to be below 1e-16.  BWP-BCH coding is also proposed for the optical transport network,
wherein only hard-decision decoding is considered \cite{ITU}.  For modern optical transport network, it is generally required that the output bit-error-rate (BER) be below 1e−15,
and the codec be able to achieve very high data rate, e.g., 100 Gbps. Thus, turbo codes and low density parity check (LDPC) codes are not suited.

In this work we systematically explore BWP-BCH codes.  Our contributions are in threefold. Firstly, we devise various new decoding algorithms for BCH codes. The proposed $-1$ decoding algorithm effectively eliminates fruitless Chien search when decoding is proclaimed unsuccessful. It is often employed in decoding of (block-wise) product BCH codes in order to reduce BCH miscorrection rate. The refined $+1$ list decoding algorithm exhibits the same complexity order as the state-of-the-art hard decoding algorithms. The proposed $+2$ list decoding algorithm exhibited a desired computational complexity of $O(n^2)$, where $n$ denotes the code length.

Secondly, we design scalable BWP-BCH codes under the given arbitrarily data and parity length, in attempt to simultaneously achieve three goals: good scalability, low encoding and decoding complexity, and good waterfall performance with low error floor. We arrange the input data in a near square array so that row and column codes share similar parameters, which allows to share the same circuit for row-wise and column-wise decoding. It also allows to accommodate slight increasing or decreasing data or parity length with minor architectural changes. We choose extended BCH (eBCH) code instead of BCH code as constituents in purpose for reducing constituent-wise miscorrection rate wile simplifying decoding process. We use an inner high-rate Reed-Solomon code to lower error floor while minimizing extra parity overhead and implementation complexity.

Lastly, we investigate efficient hard decoding of the proposed BWP-BCH codes. We aim to minimize three different error events which cause hard decoding failure. The first one is excessive errors in a few data blocks which causes both row and column decoding failure. The second one is excessive errors in a few eBCH parities, which are not cross protected. The last one is miscorrection of eBCH constituents, due to small minimum distance. We present a novel iterative decoding algorithm which is divided into three phases. The first phase iteratively applies reduced BCH correction capabilities to correct lightly corrupted rows/columns while suppressing miscorrection, until the process stalls. The second phase iteratively decodes up to the designed correction capabilities, until the process stalls. The last phase iteratively applies the proposed list decoding in a novel manner which effectively determines the correct candidate as follows. Upon successfully determining a list of candidates from a failed row (column) constituent word, trial-correction is performed on each candidate. Each time check if the crossing column (row) Berlekamp decoding of any previous failed word is successful. We choose the one that results in the most number of crossing column (row) corrections and make both row and column corrections accordingly.

The paper is organized as following. Section II devises BCH decoding algorithms for reduced-1-bit decoding, extra-1-bit decoding, and extra-2-bit decoding. Section III presents a systematic BPC-BCH code construction that carefully takes into account for performance, implementation and scalability. Section IV describes a novel iterative decoding algorithm for BWP-BCH codes aiming to improve both waterfall and error-floor performance. Section~V describes the evaluation setup and results to validate the proposed decoding algorithm. Section~VI concludes with pertinent remarks.

\section{BCH Decoding Algorithms}
\label{sec:list}

Let $t$ denote the designed error-correction capability of a (possibly
shortened) BCH ($n$, $k$) code, defined over a binary extension field
$\GF(q=2^m)$. Let $\alpha$ denote a primitive element of $\GF(q)$. The
underlying generator polynomial $g(x)$ of the BCH code contains
consecutive roots $\alpha, \alpha^2, \dots, \alpha^{2t}$, such that
\begin{equation}
g(x)\eqdef \text{LCM}\big(\mu_1(x), \mu_3(x),\ldots, \mu_{2t-1}(x)\big)
\end{equation}
where $\mu_i(x)$ denotes the minimal binary polynomial of $\alpha^i$ and LCM stands for the least common multiple (cf \cite{MacWilliams}). A counterpart extended BCH (eBCH) code adds an extra root 1, i.e.,
\begin{equation}
{\bar g}(x)\eqdef (x-1)\text{LCM}\big(\mu_1(x), \mu_3(x),\ldots, \mu_{2t-1}(x)\big).
\end{equation}
Clearly, each eBCH codeword has even Hamming weight.   It is shown that (p. 263, \cite{MacWilliams})
\begin{lemma}
The error-correction capability $t$ of a BCH $(n, k)$ code satisfies
\begin{equation}
n-k = mt
\end{equation}
and
\begin{equation}
g(x) = \mu_1(x)\mu_3(x)\ldots \mu_{2t-1}(x)
\end{equation}
when it holds
\begin{equation}
t\leq 2^{\lceil m/2\rceil -1}. \label{linear-t-condition}
\end{equation}
\end{lemma}
Note the BCH codes of practical interests have high code rates, hence, we shall practically treat $n-k = mt$ in the subsequent analyses.

Define the support of a codeword to be the index set of its nonzero entries. Next lemma sheds some light on the codeword distribution
\begin{lemma}  \label{LEM-BCH-min-support}
There does not exist a nonzero codeword whose support lies in the interval of $n-k$.
\end{lemma}
{\em Proof: } Assume otherwise is true. Let $c(x)$ be such a codeword polynomial. Then, $c(x)$ must be in the form of
$$c(x)=x^i c^*(x)$$
 where $c^*(x)$ has degree less $n-k$ by assumption. By definition, $c(x)$ divides the generator polynomial $g(x)$. Note $x^i$ is coprime with $g(x)$, thus $c^*(x)$ must divide $g(x)$. This is apparently false, since $\deg(c^*(x))<n-k=\deg(g(x))$. This concludes the lemma. \hfill $\Box\Box$

Let a (possibly shortened) BCH ($n$, $k$) code be with the designed error correction capability $t$. For a binary senseword $r(x)=\sum_{j=0}^{n-1} r_jx^j$, its syndromes are defined
\begin{equation}
S_i\eqdef r(\alpha^{i+1}), i=0, 1, \ldots, 2t-1.
\end{equation}
 The Berlekamp algorithm is a simplified version of the Berlekamp-Massey algorithm for decoding binary BCH codes
by incorporating the special syndrome property
 \begin{equation}
 S_{2i+1}=S_{i}^2, \hspace{0.2in} i=0, 1, 2, \ldots           \label{square-syndrome}
 \end{equation}
 which yields zero discrepancies at even iterations of the Berlekamp-Massey algorithm (cf. \cite{Blahut}).
Below is a concisely reformulated Berlekamp algorithm.

\vspace{0.1in}
{\fbox {\bf ALG-1: Reformulated Berlekamp Algorithm}}
{\small \begin{itemize}
\item Input: \ ${\bf S}=[S_0, \;\; S_1,\; \;S_2,\; \;\ldots,\;\; S_{2t-1}]$
\item Initialization: \ $\Lambda^{(0)}(x)=1$, $\BB^{(-1)}(x)=x$,  $L^{(0)}_\Lambda=0$, $L^{(-1)}_\BB=1$
\item For $r=0$, 2, \ldots, $2t-2$, \ do:
\begin{itemize}
\item Compute $\Delta^{(r+2)}=\sum_{i=0}^{L^{(r)}_\Lambda} \Lambda^{(r)}_i \cdot S_{r-i}$
\item Compute $\Lambda^{(r+2)}(x)=\Lambda^{(r)}(x) -\Delta^{(r+2)} \cdot \BB^{(r-1)}(x)$
\item If $\Delta^{(r+2)} \ne 0$ and $2L^{(r)}_\Lambda\leq r$, \ then
\begin{itemize}
\item Set $\BB^{(r+1)}(x)\gets (\Delta^{(r+2)})^{-1} \cdot x^2\Lambda^{(r)}(x)$
\item Set $L^{(r+2)}_\Lambda\gets L^{(r-1)}_\BB$, \ \ $L^{(r+1)}_\BB\gets L^{(r)}_\Lambda+2$
\end{itemize}
\item Else
\begin{itemize}
\item Set $\BB^{(r+1)}(x) \gets  x^2 \BB^{(r-1)}(x)$
\item Set $L^{(r+1)}_\BB\gets L^{(r-1)}_\BB+2$, \ \ $L^{(r+2)}_\Lambda\gets L^{(r)}_\Lambda$
\end{itemize}
\end{itemize}
\item Output: \ $\Lambda(x)$, \ $\BB(x)$, \ $L_\Lambda$, \ $L_\BB$
\end{itemize}
}

In the above algorithm, $\BB(x)$ is a shifted $B(x)$ which is widely used in textbooks (cf. \cite{Blahut}), $\BB(x)\eqdef x^2 B(x)$ (we found $\BB(x)$ more concise in subsequent algorithmic descriptions). It is easily observed from the above algorithm
\begin{equation}
L_\Lambda + L_\BB = 2t+1.
\end{equation}
For the conventional decoding,  the so-called Chien search (P. 164, \cite{Blahut}) is an exhaustive root search among $\{\alpha^{-i}\}_{i=0}^{n-1}$ carried out on $\Lambda(x)$. It is worth noting that, for practical high-rate BCH codes, the Chien search is far more computationally intensive than the Berlekamp algorithm and syndrome computation.  If the number of distinct roots equals to $L_\Lambda$, then all root indexes correspond to the error locations, otherwise the decoding is declared failure.

We refer ``$-1$ decoding'' to correcting up to $t-1$ errors under the designed correction capability $t$, In \cite{Al09}, it is shown that performing reduced-1-bit decoding during the first iteration  effectively achieves superior decoding performance by reducing constituent-wise miscorrection rate. In Section IV, we shall also incorporate reduced-1-bit decoding into our proposed iterative decoding of BWP-BCH codes.  In the following we present an efficient reduced-1-bit decoding algorithm.

\vspace{0.1in}
{\fbox {\bf ALG-2: $-1$ Decoding Algorithm}}
\begin{itemize}
\item Input: \ $S_0$, $S_1$, \ldots, $S_{2t-1}$
\item Apply the Berlekamp algorithm to produce $\Lambda(x)$ and $L_\Lambda$.
\item If $L_\Lambda \geq t$, then declare failure.
\item Perform the Chien search to determine all roots. If the number of distinct roots equals to $L_\Lambda$ then correct all erroneous bits, else declare failure.
\end{itemize}
Note the extra syndrome $S_{2t-1}$ is used for $t-1$-correction. Its advantage is twofold. When $t-1$ correctable, it guarantees $L_\Lambda<t$, otherwise, $L_\Lambda<t$ occurs with probability $q^{-1}$, where $q$ denotes its operation field \cite{Blahut, Wu08}, thus precluding the fruitless Chien search. Clearly, when there are $t$ errors, the Berlekamp algorithm results in $L_\Lambda=t$ and thus the Chien search is precluded.

We next introduce the list decoding algorithms to correct extra 1 bit beyond $t$.
Define
\begin{equation}
Q(x)\eqdef \frac{\Lambda(x)}{\BB(x)}  \label{def-Delta}
\end{equation}
and
\begin{equation}
Q_i = \{ j: \; Q(\alpha^{-j}) = i \}.
\end{equation}
An efficient extra-1-bit list decoding algorithm was given below, with minor modifications from \cite{Wu08}.

\vspace{0.1in}
{\fbox {\bf ALG-3: $+1$ List Decoding Algorithm}}
\begin{enumerate}
\item[Input:] \ $\Lambda(x)$, \ $\BB(x)$, \ $L_\Lambda$
\item If $L_\Lambda > t+1 $, then declare a decoding  failure.
\item If $L_\Lambda \leq t $, then determine all distinct roots in $\{\alpha^{-i}\}_{i=0}^{n-1}$.
If the number of (distinct) roots is equal to $L_\Lambda$, then return the corresponding unique codeword,
otherwise, if $L_\Lambda < t$, declare a decoding failure (which is identical to the normal Berlekamp algorithm)
\item Initialize $\delta_i=\emptyset$, $i=0$, 1, 2, \ldots, $q-1$
\item For $i=0$, 1, 2, \ldots, $n-1$, {\bf do}:
\begin{itemize}
\item Evaluate  $Q_i=\frac{\Lambda(\alpha^{-i})}{\BB(\alpha^{-i})}$.
\item If $Q_i\ne \infty$, then set $\delta_{Q_i}\gets \delta_{Q_i}  \cup \{i\}$.
\item If $|\delta_{Q_i}| = t+1$, then flip bits on indices in $\delta_{Q_i}$ and output the resulting candidate codeword.
\end{itemize}
\end{enumerate}

Note a minor correction in Step 2 is made over the original algorithm in \cite{Wu08}. Specifically, the clause ``if $L_\Lambda < t$'' is necessary for an early termination, whereas $L_\Lambda = t$ may yield valid $t+1$ error corrections.  On another note, $\BB_i=0$ results in $Q_i=\infty$. Since $\BB(x)$ is not a valid error locator polynomial so its roots are safely ignored. We observe that $\{Q_i\}_{i=0}^{q-1}$ may be efficiently implemented by link list structure at the space complexity of $O(q)$. Overall, the computational complexity of the above algorithm remains the same as the Berlekamp algorithm, i.e., $O(tn)$, but utilizing a larger space complexity of $O(q)$.

Note that $n$ terms of $\{Q_i\}_{i=0}^{n-1}$ may contribute to at most $\lfloor\frac{n}{t+1}\rfloor$ groups of $t+1$ identical values. Therefore, the above one-step-ahead algorithm may produce up to $\lfloor\frac{n}{t+1}\rfloor$ candidate codewords. An alternative interpretation is to flip each of $n$ bits and each time to apply the Berlekamp algorithm. This produces at most $n$ codewords (assuming each decoding trail is successful). Note any codeword is repeated in $t+1$ times, i.e., the same codeword is yielded by flipping any of $t+1$ error bits. However, if the actual minimum distance is at least $2t+3$, particularly for shortened codes, then there is up to single candidate.

For some (particularly low rate) BCH codes, it occurs that $S_{2t+2}$, \ldots, $S_{2t+2\tau}$ ($\tau\geq 1$) are known. By sweeping $S_{2t}$ over $\GF(q)$, up to $t+1+\tau$ errors are list decoded with computational complexity of $O(qnt)$. The algorithm is detailed below

\vspace{0.1in}
{\fbox {\bf ALG-4: $+\tau+1$ List Decoding with Known $\{S_{2t+2i}\}_{i=1}^\tau$ }}
\begin{itemize}
\item[Input:] \ $\Lambda^{(2t)}(x)$, \ $\BB^{(2t-1)}(x)$, \ $L^{(2t)}_\Lambda$, \ $L^{(2t-1)}_\BB$, \ $\{S_{2t+2i}\}_{i=1}^\tau$
\item For $S_{2t}=0$, 1, 2, \ldots, $q-1$, \ {\bf do}:
  \begin{itemize}
  \item For $r=2t$, $2t+2$, \ldots, $2t+2\tau$, \ {\bf do}:
      \begin{itemize}
      \item Compute $\Delta^{(r+2)}=\sum_{i=0}^{L^{(r)}_\Lambda} \Lambda^{(r)}_i \cdot S_{r-i}$
      \item Compute $\Lambda^{(r+2)}(x)=\Lambda^{(r)}(x) -\Delta^{(r+2)} \cdot \BB^{(r-1)}(x)$
      \item If $\Delta^{(r+2)} \ne 0$ and $2L^{(r)}_\Lambda\leq r$, \ then
           \begin{itemize}
           \item Set $\BB^{(r+1)}(x)\gets (\Delta^{(r+2)})^{-1} \cdot x^2\Lambda^{(r)}(x)$
           \item Set $L^{(r+2)}_\Lambda\gets L^{(r-1)}_\BB$, \ \ $L^{(r+1)}_\BB\gets L^{(r)}_\Lambda+2$
           \end{itemize}
      \item Else
           \begin{itemize}
           \item Set $\BB^{(r+1)}(x) \gets  x^2 \BB^{(r-1)}(x)$
           \item Set $L^{(r+1)}_\BB\gets L^{(r-1)}_\BB+2$, \ \ $L^{(r+2)}_\Lambda\gets L^{(r)}_\Lambda$
      \end{itemize}
   \end{itemize}
\item Perform the Chien search on $\Lambda(x)$. If the number of distinct roots equals to $L_\Lambda$, then output the resulting candidate codeword.
\end{itemize}
\end{itemize}

Consider the (63, 24) BCH code with $t=7$. Its generator polynomial contains the roots $\alpha$, $\alpha^3$, \ldots, $\alpha^{13}$,  but also two extra roots $\alpha^{17}$ and $\alpha^{19}$. For this code, up to 3 extra errors, i.e., up to 10 errors, can be listed decoded by sweeping $S_{14}$. Note the above algorithm may further incorporate the preceding extra-1-bit decoding to achieve extra-$\tau+2$-bit list decoding at the same order of complexity.

We proceed to present an efficient extra-2-bit list decoding algorithm. The basic idea is to apply a one-pass Chase decoding \cite{Wu12} and to follow with the above extra-1-bit decoding. \cite{Wu12} described a one-pass Chase decoding algorithm in which the error locator polynomial associated with flipping a bit can be obtained in constant time and with the computational complexity of $O(t)$. The following describes one-pass one-bit flipping Chase decoding.

\vspace{0.1in}
{\fbox {\bf ALG-5: One-Pass One-Bit Flipping Chase Decoding Algorithm}}
\begin{itemize}
\item[Input:] \ $\Lambda(x)$, \ $\BB(x)$, \ $L_\Lambda$, \ $L_\BB$
\item For $i=0$, 1, 2, \ldots, $n-1$, \ do:
{\small

\begin{enumerate}

\item Evaluate \ $\Lambda_i\gets \Lambda(\alpha^{-i}), \hsp \BB_i\gets \BB(\alpha^{-i})$

\item Update polynomials:

\begin{itemize}

\item {\bf Case 1}: $\Lambda_i=0$  $\vee$ ($\Lambda_i\ne 0$ $\wedge$ $\BB_i\ne 0$ $\wedge$ $L_\Lambda \geq L_\BB$)
\begin{equation*}
\left\{ \begin{array}{l}
\Lambda^{(i)}(x) \gets  \BB_i \cdot \Lambda(x) - \Lambda_i \cdot \BB(x) \\
\BB^{(i)}(x)  \gets   (x^2-\alpha^{-2i}) \BB(x) \\
L^{(i)}_\Lambda \gets  L_\Lambda,  \hsp        L^{(i)}_\BB \gets  L_\BB+2
\end{array}\right.
\end{equation*}

\item {\bf Case 2}: $\BB_i=0$  $\vee$ ($\Lambda_i\ne 0$ $\wedge$ $\BB_i\ne 0$ $\wedge$ $L_\Lambda<L_\BB-1$)
\begin{equation*}
\left\{ \begin{array}{l}
\Lambda^{(i)}(x) \gets  (x^2-\alpha_i^{-2}) \Lambda(x)  \\
\BB^{(i)}(x)  \gets  \BB_i \cdot x^2\Lambda(x) - \alpha^{-2i}\Lambda_i \cdot \BB(x) \\
L^{(i)}_\Lambda \gets  L_\Lambda+2, \hsp        L^{(i)}_\BB \gets  L_\BB
\end{array}\right.
\end{equation*}

\item {\bf Case 3}: $\Lambda_i\ne 0$ $\wedge$ $\BB_i\ne 0$ $\wedge$ $L_\Lambda=L_\BB-1$
\begin{equation*}
\left\{ \begin{array}{l}
\Lambda^{(i)}(x) \gets \BB_i \cdot \Lambda(x) - \Lambda_i \cdot \BB(x)  \\
\BB^{(i)}(x)  \gets  \BB_i \cdot x^2\Lambda(x) - \alpha^{-2i}\Lambda_i \cdot \BB(x) \\
L^{(i)}_\Lambda \gets  L_\Lambda+1, \hsp        L^{(i)}_\BB \gets  L_\BB+1
\end{array}\right.
\end{equation*}

\end{itemize}

\end{enumerate} }

\end{itemize}

For each pair $\left(\Lambda^{(i)}(x), \BB^{(i)}(x)\right)$, $i=0, 1, \ldots, n-1$, we may apply the proposed $+1$ decoding algorithm to determine all candidates up to $t+1$ bits difference (note the index $i$ is pre-flipped). Thus, combining the above one-pass one-bit flipping Chase decoding algorithm and the $+1$ decoding algorithm effectively list decodes all codewords up to distance $t+2$. Clearly, the overall computational complexity is $O(n^2t)$, due to $n$ deployments of extra-1-bit decoding. We now explore ways to reduce complexity. We first note that $\{(\Lambda_i,\; \BB_i)\}_{i=0}^{n-1}$ are evaluated by the above algorithm, so, instead of updating polynomial pairs $(\Lambda^{(i)}(x), \BB^{(i)}(x))$, $i=0, 1, \ldots, n-1$, it takes $O(n)$ to evaluate each vector pairs $\left( \{\Lambda_j^{(i)} \}_{j=0}^{n-1},\;\; \{\BB_j^{(i)} \}_{j=0}^{n-1} \right)$ for an index $i$. Consequently, the $+1$ decoding algorithm takes merely $O(n)$ complexity. We further note that, when $i$-th index is flipped for $+1$ decoding, all candidates of $t+2$ bits correction involving at least a bit among indexes \{0, 1, 2, \ldots, $i-1\}$ have been listed, therefore, the extra-1-bit decoding algorithm associated with $i$-th bit flipping suffices to search through the index subset $\{i+1, i+2, \ldots, n-1\}$.  The detailed algorithmic procedure is described below.

\vspace{0.1in}
\noindent
{\fbox {\bf ALG-6: $+2$ List Decoding Algorithm}}
\begin{enumerate}
\item[Input:] \ $\Lambda(x)$, \ $\BB(x)$, \ $L_\Lambda$, \ $L_\BB$

\item Evaluate and store \  $\{\Lambda_i\}_{i=0}^{n-1} \gets \{\Lambda(\alpha^{-i})\}_{i=0}^{n-1}, \hsp \{\BB_i\}_{i=0}^{n-1} \gets \{\BB(\alpha^{-i})\}_{i=0}^{n-1}.$

\item {\bf For} $i=0$, 1, 2, \ldots, $n-t-2$, \ {\bf do}:
{\small

\begin{enumerate}

\item Initialize $\delta_j=\emptyset$, $j=0$, 1, 2, \ldots, $q-1$.
\item For $j= i+1$, $i+2$, \ldots, $n-1$, \ {\bf do}:
\begin{itemize}

\item Compute
\begin{equation*}
{\bar Q}_j  \gets
\begin{cases}
\frac{  \BB_i \Lambda_j + \Lambda_i  \BB_j }{ (\alpha^{-2(j-i)}+1)\BB_j }, & \text{if } \Lambda_i=0 \vee  (\Lambda_i\ne 0 \wedge \BB_i\ne 0  \wedge L_\Lambda \geq L_\BB) \\
\frac{ (\alpha^{-2j} + \alpha^{-2i}) \Lambda_j }{  \alpha^{-2(j-i)} \BB_i \Lambda_j+\Lambda_i \BB_j }, & \text{if } \BB_i=0 \vee (\Lambda_i\ne 0 \wedge \BB_i\ne 0 \wedge L_\Lambda<L_\BB-1) \\
\frac{ \BB_i \Lambda_j +  \Lambda_i \BB_j }{ \alpha^{-2(j-i)} \BB_i \Lambda_j +  \Lambda_i  \BB_j }, & \text{otherwise}
\end{cases}
\end{equation*}

\item If ${\bar Q}_j\ne \infty$, then set $\delta_{{\bar Q}_j}\gets \delta_{{\bar Q}_j}  \cup \{j\}$.
\item If $|\delta_{{\bar Q}_j}| = t+1$, then flip bits on indices in $\delta_{{\bar Q}_j}\cup \{i\}$ and output the resulting candidate codeword.

\end{itemize}
\end{enumerate} }

\end{enumerate}

Note that ${\bar Q}_j$ in the above Step 2.b is scaled by a constant $\alpha^{-2i}$  without altering result.  Clearly, the above algorithm exhibits a computational complexity of $O(n^2)$ and space complexity of $O(q)$. Assume in a perfect scenario that flipping each of two bits results in a candidate with $t$ errors. There are up to $\binom{n}{2}$ candidate codewords. Note each candidate is exactly repeatedly counted in $\binom{t+2}{2}$ times, this is because flipping any of 2 out of $t+2$ errors is corrected to the same codeword.   Therefore, the number of candidate codewords is bounded by $\frac{n(n-1)}{(t+2)(t+1)}$. However, if the actual minimum distance is at least $2t+3$, particularly in the case of shortened codes, then there exist up to $\lfloor \frac{n}{t+2} \rfloor$ candidates. Moreover, if the actual minimum distance is at least $2t+5$, particularly in the case of highly shortened codes, then there is up to single candidate.

An alternative but less efficient approach to perform extra-2-bit list decoding is by sweeping all possibilities of $S_{2t}$, equivalently all possibilities of $\Delta_{2t+2}$, and then deploying extra-1-bit list decoding over each of $q$ pairs of $\big(\Lambda(x), \; \BB(x)\big)$. Its complexity, after appropriate optimization, is reduced to $O(qn)$. This approach is akin to the $t+1$ list decoding algorithm for Reed-Solomon codes \cite{Egorov04}. The proposed $+2$ decoding algorithm is clearly advantageous when $n\ll q$, which often holds true during iterative decoding of BPC-BCH and other product-like BCH codes.

Algorithms presented in this section assume the consecutive error locators, $\{\alpha^i\}_{i=0}^{n-1}$, as defined in conventionally shortened codes. In the scenario of BWP-BCH decoding, due to partial correction by the cross decoding, the error locators are usually not consecutive. Let $\{\alpha_i\}_{i=0}^{n^*-1}$ denote the set of uncorrected error locators (where $n^*\leq n$). To this end, $\alpha^i$ (likewise $\alpha^j$) in the above algorithms are to be replaced by $\alpha_i$, such that, $\alpha^{-i}\to \alpha_i^{-1}$, $\alpha^{-2i}\to \alpha_i^{-2}$.

\begin{figure}[t]
\begin{center}
\centering \includegraphics[width=6.8in]{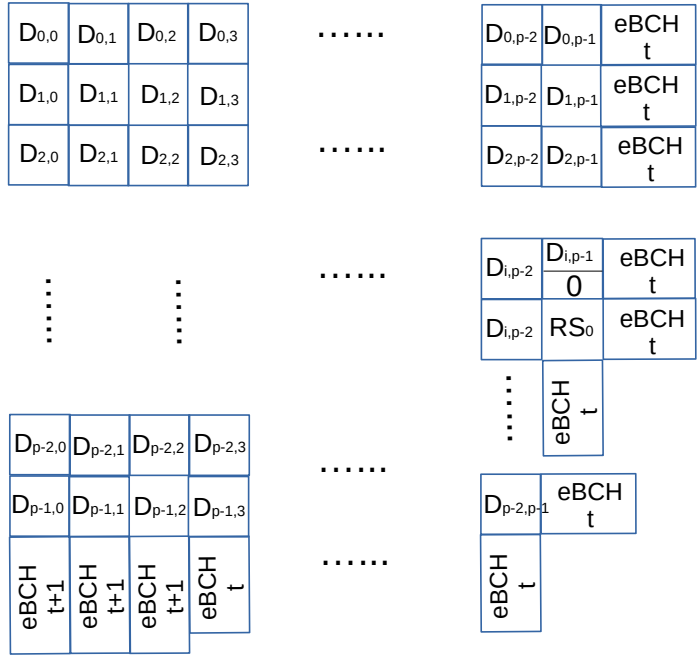}

\vspace{-0.5in}
\caption{An example illustration of the case-1 BWP-BCH codeword.} \label{FIG:codeword-case1}
\end{center}
\end{figure}

\begin{figure}[t]
\begin{center}
\centering \includegraphics[width=6.8in]{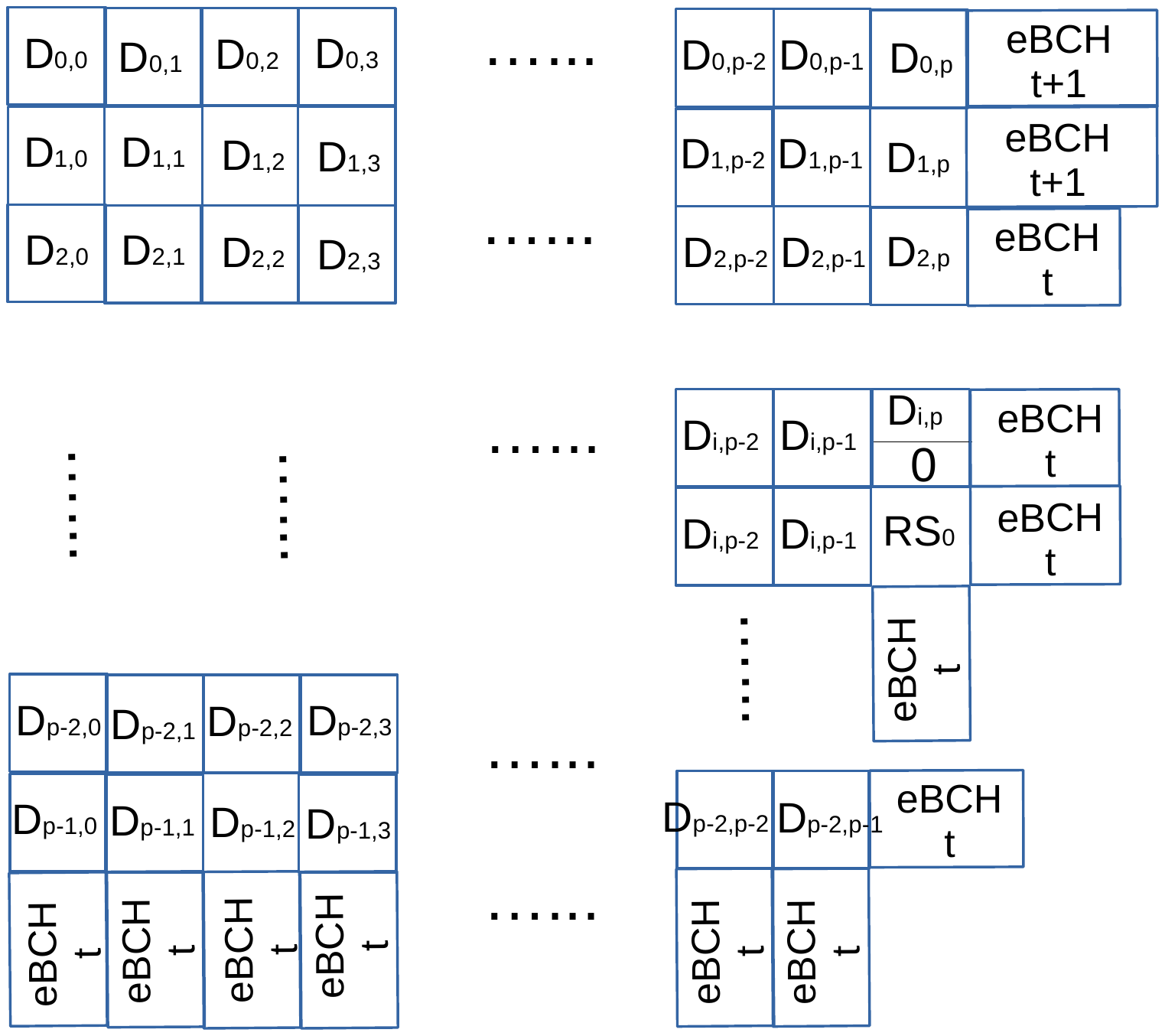}

\caption{An example illustration of the case-2 BWP-BCH codeword.} \label{FIG:codeword-case2}
\end{center}
\end{figure}

\section{Designing Scalable Block-Wise Product BCH Codes}

Instead of following conventional wisdom of studying a regularly defined BWP-BCH code, we start from scratch with an arbitrarily given message length $K$ and parity length $R$. We leverage the following freedoms to design a ``good'' BWP-BCH code, wherein ``good'' qualitatively means good waterfall performance and low error floor, \\
$(i)$.  block size $b$;   \\
$(ii)$. BCH vs. eBCH;\\
$(iii)$. serial vs. parallel concatenation; \\
$(iv)$. square vs. rectangular shape in organizing message blocks; \\
$(v)$.  concatenation of an outer or inner erasure code.

Specifically, we leverage these freedoms under the following guidelines.\\
$(i)$. Block size $b$ affects the BCH message length, operation field, and its error correction capability. $b$ must be relatively large so that the resulting BCH codes exhibit low miscorrection rate. It is preferred to have $t\geq 4$.  Also, it is preferred to choose $b$ a multiple of 8 to facilitate hardware implementation. However, It is unnecessary to force $b$ dividing $K$, as opposed to the literature \cite{Cho14, Kim15}. This is because a partial last message block (wherein $b$ does not divide $K$) can be mitigated at no rate penalty. Specifically, the partial block is padded with $\left\lceil \frac{K}{b} \right\rceil \cdot b -K$ zeros to form a full block for both inner and outer coding, but the padded zeros are not actually transmitted and the equal number of zeros are re-padded at the receiver. Though it is possible to harness the error-free property of the padded zeros to rule out certain miscorrection, we suggest to treat the padded block as a regular block, so that all row and column BCH constituent words are treated unanimously.   \\
$(ii)$. An eBCH code includes an extra parity bit to enforce even code weights, and thus halves the miscorrection rate. Using a parity bit also allows to reduce decoding complexity. For a given eBCH senseword, either $+1$ list decoding or $+2$ list decoding is applicable, but not both. This is because, the parity syndrome being 0 indicates that the number of errors must be even, whereas 1 indicates an odd number of errors.  In \cite{Chen09}, it is proven that the number of Chase-II decoding trials of eBCH codes can be cut by half without performance loss, rendering more efficient soft-decision decoding of product eBCH codes. Our extensive simulations also indicate that BWP-eBCH codes performs slightly superior to the counterpart BWP-BCH codes, attributed to the smaller miscorrection of eBCH codes. Therefore, we shall use eBCH, instead of BCH, codes as constituent code. \\
$(iii)$. In parallel concatenation, each message block is protected by a row and column eBCH code, but neither row nor column eBCH parity is protected by the other, whereas in serial concatenation, there is one dimension of parities that are covered by the other dimension of parities. It is known that BWP codes loose the key property of parity-on-parity for conventional product codes. Parity-on-parity in conventional product codes yields a minimum distance of product of row and column minimum distances but it is not true for BWP codes even with protecting one-dimensional parity, i.e., serial concatenation. In short, serial concatenation does not exhibit a conspicuous property, such as a product of minimum distances.  Therefore, we choose parallel concatenation wherein row and column decoding are symmetrical. The symmetry effectively enables to design the same circuit for row and column decoding. Moreover, parallel concatenation inherently allows for row and column encoding in parallel. \\
Lemma~\ref{LEM-BCH-min-support} sheds some light on minimum weight codewords (associated with the minimum distance). There is no codeword whose support lies only in eBCH parities. Furthermore, if the block size is no greater than eBCH parity length (which is often the true case), then there does not exist a codeword whose support lies only in one message block. An inner $f$-erasure RS code guarantees at least $f+1$ non-zero blocks, yielding at least $f+1$ uncorrelated non-zero eBCH codewords. \\
$(iv)$. Message blocks are organized in a near square shape so that row and column decoding operations are nearly identical. To enforce code scalability as well as implementation simplicity, all eBCH codes are defined in the same field (even if the last data column has a single block).\\
$(v)$. Concatenation of an erasure code mitigates error floor at the price of lower BCH correction capabilities. An inner Reed-Solomon (RS) code is incorporated to mitigate error floor. This is different from \cite{Yu14}, wherein an outer erasure code is concatenated. An outer code has to protect both message blocks, eBCH parities, along with its own parity, therefore, it demands for more parity than an inner code which just protects message blocks. The rationale behind not protecting eBCH parities is that, when rows/columns have few errors in their data messages but excessive errors in parities which result in decoding failure, the errors in data messages can be corrected by cross decoding predominantly, thus the errors in parities can be simply recovered through re-encoding.  In fact,  our extensive simulations indicate that up to 4 RS parity symbols suffice to reduce error floor satisfactorily for our codes of interest. Therefore, adopting an inner RS code reserves more redundancy for eBCH parity but also reduces its implementation complexity. Note that two erasure blocks enables to recover single-row plus two-column failures or single-column plus two-row failures. Likewise,  three erasure blocks enables to recover single-row plus three-column failures or single-column plus three-row failures. Four erasure blocks enables to recover  single-row plus four-column failures, single-column plus four-row, or two-row plus two-column failures.   When inner code is not used, i.e., $f=0$, the BWP-BCH code suffers from the dominant failure mechanism of single uncorrectable row and column \cite{Kim15}.  The authors in \cite{Kim15} proposed a collaborative method to forge a new BCH senseword from the failed row and column constituents. However, this approach only succeeds in some cases, while taking extra endeavor to forge new syndromes. On the other hand, the single row and column failure is handily thwarted using a parity inner coding, at the cost of one parity block overhead.   \\
$(vi)$. It allows to easily accommodate multiple message lengths and parity lengths in high granularity. This is particularly important in data storage, wherein different vendors/customers have slightly different requests.

Given the block size $b$, the number of message blocks is $\lceil \frac{K}{b} \rceil$.  Assume also $f$-erasure RS encoding is deployed. The inner RS code length is given by
$\eta \eqdef \lceil \frac{K}{b} \rceil + f$. Evidently, the minimum field dimension for RS coding is  $\lceil \log_2 \eta \rceil$. Let $p$ satisfies
\begin{equation}
p(p-1) < \eta \leq p(p+1).   \label{block-square-dimension}
\end{equation}
Note the positive number set, $\mathbb{Z}^+$, is disjointedly partitioned into
$$\mathbb{Z}^+=\sum_{p=1}^\infty \Big[(p(p-1), \;\; p(p+1)\Big].$$
Thus, any positive number is uniquely distributed to one of intervals $\big(p(p-1),\; p(p+1)\big]$.
To solve for $p$, first let $a$ be the real positive root of the equation
$$a(a+1)=\eta.$$
We obtain
$$a=\frac{-1+\sqrt{1+4\eta}}{2}.$$
It is easily verified that $p=\lceil a \rceil$, i.e.,
\begin{equation}
p = \left\lceil \frac{-1+\sqrt{1+4\eta}}{2} \right\rceil
\end{equation}
is the unique solution of \eqref{block-square-dimension}.

We further partition into two cases. The first case is such that
\begin{equation}
p(p-1) < \eta \leq p^2.
\end{equation}
Then the inner RS code is arranged into $p\times p$ matrix. There are $2p$ eBCH words, each is allocated with average $\lceil\frac{R-fb}{2p} \rceil$ parity bits. Accordingly, the eBCH code field dimension is determined by
\begin{equation}
m = \left \lceil \log_2 \left(pb +  \left\lceil\frac{R-fb}{2p} \right\rceil \right) \right \rceil,  \label{field-dimension}
\end{equation}
and the base correction capability is given by
\begin{equation}
t = \left \lfloor \frac{R-fb-2p}{2pm} \right\rfloor,
\end{equation}
wherein we assume the code rate is high enough to meet the condition of Lemma 1. For lower rate codes where Lemma 1 does not apply, $t$ is determined through computer search.
Note there remains extra correction power of
\begin{equation}
\theta = \left \lfloor \frac{R-fb-2p}{m} \right \rfloor - 2pt.
\end{equation}
When $\theta>0$, the $\tau$ longest inner block rows/columns (in the sequel, a row/column always implies a block row/column) are assigned with $t+1$ correction capability, whereas the remaining $2p-\theta$ inner blocks rows/columns are assigned with $t$ correction capability. In this case, due to uneven distribution of eBCH parities, it is necessary to cross check the validity of \eqref{field-dimension}, such that
\begin{equation}
pb + (t+1)m+1 < 2^m.
\end{equation}
Figure~\ref{FIG:codeword-case1} illustrates an example of the above BWP-BCH code description, wherein the last partial data block is padded with zeros and single-parity RS code is used.

The second case is such that
\begin{equation}
p^2 < \eta \leq p(p+1).
\end{equation}
Then the inner RS code is arranged in a $p\times (p+1)$ matrix. There are $2p+1$ eBCH words, each is allocated with average $\lceil\frac{R-fb}{2p+1} \rceil$ parity bits. Accordingly, the eBCH code field dimension is determined by
\begin{equation}
m = \left \lceil \log_2 \left((p+1)b +  \left\lceil\frac{R-fb}{2p+1} \right\rceil \right) \right \rceil,
\end{equation}
and the base correction capability is given by
\begin{equation}
t = \left \lfloor \frac{R-fb-(2p+1)}{(2p+1)m} \right \rfloor.
\end{equation}
The residual correction power is determined by
\begin{equation}
\theta = \left\lfloor \frac{R-fb-(2p+1)}{m} \right\rfloor - (2p+1)t.
\end{equation}
Likewise, the $\theta$ longest inner block rows/columns are assigned with $t+1$ correction capability, whereas the remaining $2p+1-\theta$ inner blocks rows/columns are assigned with $t$ correction capability. When $\theta>0$, it is necessary to validate the field dimension $m$ such that
\begin{equation}
(p+1)b + (t+1)m+1 < 2^m.
\end{equation}
Figure~\ref{FIG:codeword-case2} illustrates an example of the above BWP-BCH code construction.

Clearly, the foregoing coding configuration is totally determined by the two parameters, $b$ and $f$. $b$ is purposed to optimize the waterfall performance. $f$ is mainly associated with error floor. The larger $f$ yields the lower error floor, however, at the price of rate penalty. Since the inner blocks are arranged in a (near) square shape, the dominant error event is such that the equal number of rows and columns are uncorrectable, yielding a square number of intersecting blocks. For this reason, it is preferred to choose $f$ to be a square number, say 1, 4. It is shown in our simulations that $f=4$ achieves good balance between low error floor and superior waterfall. The failure probability of $i$ rows failures and $j$ columns failures are extensively investigated in literature (\cite{Cho14, Yu14, Kim15}). Our case also needs to take into account for different correction capabilities among rows (columns).

Consider a data storage example wherein the data length is 4K bytes, i.e., $K=32768$, and the parity length is 455 bytes, i.e., $R=3640$. The code rate is 0.9. Assume block size to be $b=32$ and RS parity length is $f=4$. The number of inner code blocks is then $\left\lceil \frac{K}{b} \right\rceil + f = 1024+4=1028$. Accordingly, it belongs to the case 2, and results in $p=32$. The inner RS code is organized into 32 rows by 33 columns, wherein the last column has only 4 blocks.  The resulting eBCH codes are defined over the field dimension of $m=11$.  The base correction capability is given by $t = \left \lfloor \frac{3640-193}{65\times 11} \right \rfloor = 4$. The residual correction power is determined by $\tau = \left\lfloor \frac{3640-193}{11} \right\rfloor - 65\times 5  = 53$. We obtain the complete BWP-BCH configuration as in Table~\ref{Tab:Codemapping1}.

\begin{table}
\centering
\caption{BWP-BCH mapping for $(K=32768, R=3640, b=32, f=4)$. \label{Tab:Codemapping1}}
\begin{tabular}{|l|c|c|c|lll}
\hline
Rows/Columns & Inner Blocks & eBCH $t$ \\ \hline
4 rows & 33 & 5 \\ \hline
28 rows & 32 & 5 \\ \hline
21 columns & 32 & 5 \\ \hline
10 columns & 32 & 4 \\ \hline
1 column  & 4   & 4 \\  \hline
\end{tabular}
\end{table}

Consider a different block size $b=15$ while keeping $f=4$. The number of inner blocks is now 2189. Accordingly, the inner RS code is organized into 47 by 47 block matrix, wherein the last column has 27 blocks. The resulting eBCH codes are defined in 10-bit field, with the base correction capability $t=\left \lfloor \frac{3640-154}{2\times 47\times 10} \right\rfloor = 3$. The residual correction power is $\tau = \left\lfloor \frac{3640-154}{10} \right\rfloor - 94\times 3  = 66$. Table~\ref{Tab:Codemapping2} shows the detailed BWP-BCH configuration.

\begin{table}
\centering
\caption{BWP-BCH mapping for $(K=32768, R=3640, b=15, f=4)$. \label{Tab:Codemapping2}}
\begin{tabular}{|l|c|c|c|lll}
\hline
Rows/Columns & Inner Blocks & eBCH $t$ \\ \hline
27 rows & 47 & 4 \\ \hline
20 rows & 46 & 4 \\ \hline
19 columns & 47 & 4 \\ \hline
27 column  & 47   & 3 \\  \hline
1 column  & 24   & 3 \\  \hline
\end{tabular}
\end{table}

\begin{figure}[h]
\begin{center}
\centering \includegraphics[width=6.8in]{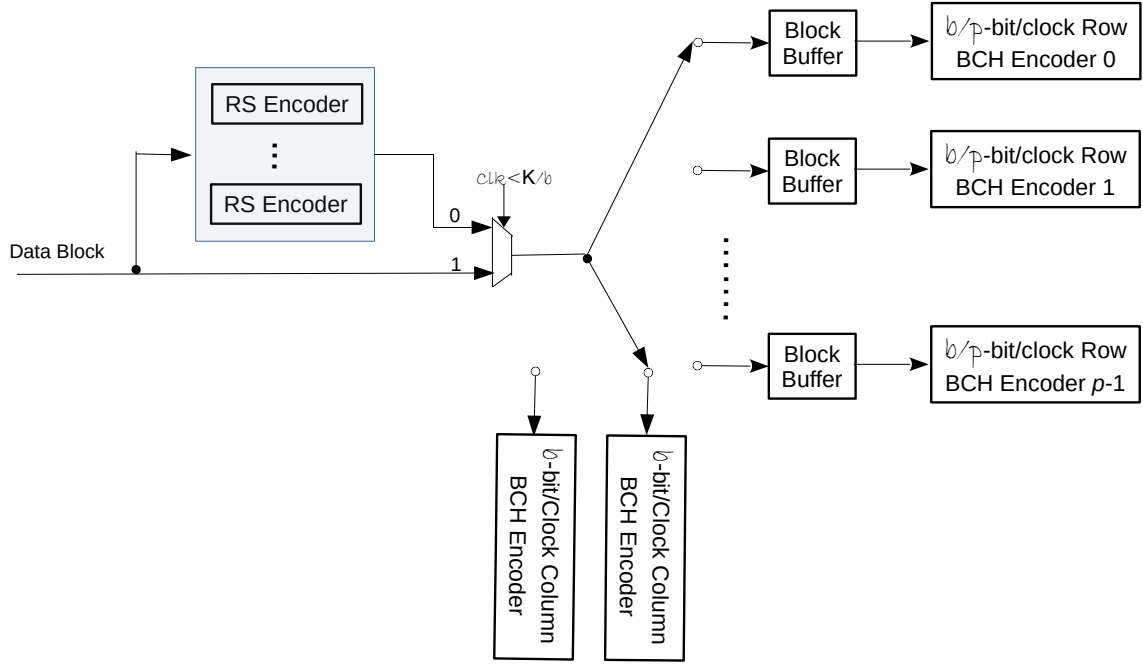}

\vspace{-0.5in}
\caption{Encoder block diagram for BWP-BCH codes} \label{FIG:Encoder}
\end{center}
\end{figure}

Note one condition must be satisfied to enforce nontrivial RS coding
\begin{equation}
2^b \geq \eta \eqdef \left\lceil \frac{K}{b} \right\rceil +f,    \label{def-eta}
\end{equation}
where $f>1$ denotes the number of RS parity blocks and $\eta$ denotes the number of inner RS blocks (for short, inner blocks). However, this enforcement is not needed if $f=1$, i.e., in the case of trivial parity coding.
For implementation simplicity, we choose its generator polynomial
\begin{equation}
g_{RS}(x) = (x-1)(x-\beta)\ldots(x-\beta^{f-1})
\end{equation}
where $\beta$ denotes a primitive element of the RS operation field. Its erasure-only decoding effectively recovers up to $f$ erased symbols, as briefly described below (cf. \cite{Blahut}).
Upon receiving a senseword $y(x)$, its syndromes are computed as follows
\begin{equation}
{\hat S}_i = y(\beta^i), \;\;\;\; i=0, 1, \ldots, f-1,
\end{equation}
wherein the notation ${\hat S}$ is to differentiate from eBCH syndromes $S$. Let $X_0$, $X_1$, \ldots, $X_{e-1}$ ($e<f$), be (known) erasure locators. Then its erasure locator polynomial is given by
\begin{equation}
{\hat \Lambda}(x) \eqdef (1-X_0x)(1-X_1x)...(1-X_{e-1}x)
\end{equation}
and erasure evaluator polynomial by
\begin{equation}
{\hat \Omega}(x) \eqdef {\hat \Lambda}(x) {\hat S}(x) \pmod{x^f}.
\end{equation}
Erasure-only decoding is successful if and only if
\begin{equation}
\deg({\hat \Omega}(x)) < e.
\end{equation}
If true, then the corresponding erasure values, $\{Y_i\}_{i=0}^{e-1}$, are retrieved by
\begin{equation}
Y_i = \frac{{\hat \Omega}(X_i^{-1})}{ {\hat \Lambda}_\text{odd}(X_i^{-1}) }, \;\;\; i=0, 1, \ldots, e-1,
\end{equation}
where ${\hat \Lambda}_\text{odd}(x)$ denotes the odd term polynomial of ${\hat \Lambda}(x)$.

In some cases \eqref{def-eta} is not met, then RS coding with $f>1$ is infeasible. We next consider to relax  \eqref{def-eta}. Let $[D_{i,j}]$ be $p\times (p+1)$ data block array, wherein empty array blocks are treated as zero blocks. The RS data vector, $[{\hat D}_0, {\hat D}_1, {\hat D}_2, \ldots, {\hat D}_{2p-2}]$, is produced by XORing $[D_{i,j}]$ reverse diagonally, i.e.,
\begin{equation}
{\hat D}_i\eqdef \oplus_{l+j=i}D_{l, j}, \;\;\; i=0, 1, \ldots, 2p-2,    \label{RS-Data-Symbol}
\end{equation}
where $\oplus$ denotes bit-wise XOR. Note $D_{2p-1}$ is not defined as $D_{p-1, p}$ is reserved for RS parity.  Accordingly, \eqref{def-eta} is relaxed to
\begin{equation}
2^b > 2p-1+f.
\end{equation}
It is worth noting that employing \eqref{RS-Data-Symbol} also renders simpler implementation. This is because a block can be partitioned into multiple sub-blocks such that each of them is separately protected by an RS code defined over a small operation field.
Clearly, failed blocks in single-column plus multiple-rows or single-row plus multiple-columns belong to different RS symbols, and thus can be uniquely recovered. However, it may not fully work for $f=4$. This is because, when two-row plus two-column failure occurs, two failed blocks may line diagonally and thus belong to the same RS symbol. In this case, the remaining two failed blocks must belong to different RS symbols and thus are uniquely recovered. Consequently, the remaining two (uncorrelated) blocks are predominantly corrected by decoding row-wise or column-wise. For implementation simplicity, it is desirable to use the above method when $f\leq 4$.

We next describe an efficient high-speed encoder. First note it is common that block size is much greater than the required RS symbol size, i.e., $ b \gg \log_2(\eta)$.
Instead of treating a block as an RS symbol, we divide a block into multiple RS symbols and encode each to a separate RS code. This way dramatically reduces the circuit complexity of finite field multiplier and divisor.
As in the above example, the 10-bit RS coding suffices for $b=40$. Thus, it suffices to partition each block into 4 RS symbols and to encode to 4 RS codes respectively.
Secondly, assume a column-wise block of data is transferred each clock, theoretically we may use one high-speed BCH encoder (cf. \cite{Pei92, Zhang05}) to process $b$ bits in a clock. However, there is difficulty to offload parity and immediately switch to  next column encoding (with proper register initializations). To this end, we use two column encoders to ping-pong for the task. On the other hand, a row encoder only needs to process a block of data upon transferring $p$ column-wise blocks of data. Our solution is to add a block buffer in front of each row encoder and design a low-performance encoder such that it processes only $\lceil b/p\rceil $ bits per cycle (eBCH encoder is halted if completed less than $p$ cycles). As far as encoding a block, row and column eBCH encoders, as well as RS encoders, follow the same first-in-first-out bit sequence. Figure~\ref{FIG:Encoder} depicts the block diagram of the proposed $b$-bits/clock encoder. Note that the proposed encoder applies for any inner block length within $\big(p(p-1),\; p(p+1)\big]$, at the same time accommodates different eBCH parities. It is worth pointing out that the enforcement of single BCH field allows to effectively share BCH encoder/decoder.

\section{New Iterative harding Decoding of BWP-BCH Codes}

Apparently, all existing BWP-BCH decoding algorithms are applicable (possibly with minor modification) to the proposed codes. In this section, we explore more efficient hard decoding of the proposed BWP-BCH codes. We aim to lower error floor by effectively handling three types of dominant error events. The first one is excessive errors in a few data blocks which causes both row and column decoding failure. The second one is excessive errors in a few eBCH parities, which are not cross protected. The last one is miscorrection of eBCH constituents, due to small minimum distance. We also aim to boost waterfall performance through intelligently incorporating the proposed extra-1-bit and extra-2-bit list decoding algorithms.
We present a novel iterative decoding algorithm for the proposed BWP-BCH codes, in the following three phases.
\begin{itemize}
\item[I.] Iteratively alternate row and column reduced-1-bit decoding until the process stalls or a pre-determined maximum number of iterations is reached.
\item[II.] Iteratively alternate row and column regular decoding until the process stalls or a pre-determined maximum number of iterations is reached.
\item[III.] Iteratively alternate row and column list decoding up to extra-2-bit errors until the process stalls or a pre-determined maximum number of iterations is reached.
\end{itemize}

The underlying purpose of Phase-I is to reduce miscorrection rate so as to avoid error amplification. We next deep dive into the implementation details.
We call {\em decoding stalling} if the numbers of failed rows and columns remain unchanged in a full iteration. Upon decoding stalling, let the number of failed intersecting blocks be the product of the number of failed rows by the number of failed columns.  Upon the completion of each half-iteration, i.e., row-wise or column-wise eBCH decoding, the decoding status is checked; the process is early terminated if a decoding success is declared. Herein the decoding success is defined as {\em the number of failed intersecting blocks is up to $f$, and RS erasure-only decoding is successful.}  We give the following examples to clarify this criterion.
\begin{itemize}
\item If all row (column) eBCH constituents are successfully decoded but not all column (row) eBCH constituents, and RS syndromes are zeros, then, the number of failed intersecting blocks is zero and thus is declared success even without inner RS coding, wherein failed constituents can be simply corrected by re-encoding.
\item If all row (column) eBCH constituents are successfully decoded but not all column (row) eBCH constituents, and RS syndromes are non-zeros, then, it is proclaimed unsuccessful.
\item If the number of failed intersecting blocks is less than $f$, then, it is declared success only if erasure-only decoding is successful, but not erasure-and-error decoding. When erasure-only decoding is successful, re-encoding is deployed to correct eBCH parities.
\item If the number of failed intersecting blocks is greater than $f$, then declare failure.
\item If all eBCH constituents are successfully decoded but RS syndromes are non-zeros, then it is proclaimed unsuccessful. This is because our designing purpose of inner RS coding is for erasure recovery, whereas random error correction may result in intractable decoding behaviors.
\end{itemize}

We next present the details of syndrome computation and update. When sequentially receiving the senseword,  the (single-bit) parity syndrome and even-indexed syndromes for each eBCH codes (note the odd-indexed syndromes are not saved but computed on-the-fly through \eqref{square-syndrome}) and  RS syndromes are simultaneously computed. Each time an eBCH constituent is successfully decoded, corrections are immediately made to the senseword, and syndromes of crossing eBCH constituents and the RS code(s) are updated accordingly.  When indexes, denoted by $\{i_l\}_{l=0}^{\iota-1}$ associated with a $t$-correcting eBCH constituent is corrected by decoding of crossing eBCH constituents, its syndromes are updated as follows.
\begin{equation}
S_j \gets S_j + \sum_{l=0}^{\iota-1} \alpha_{i_l}^{j+1}, \;\;\;\; j=0, 2, \ldots, 2t-2.
\end{equation}
It is worth noting that the parity syndrome is used to eliminate unnecessary decodings. Specifically, if the parity syndrome plus the targeted number of corrections is an odd number then the corresponding decoding is deemed unsuccessful.

In Phase-II, the decoding is limited to failed rows/columns. Upon stalling, if the number of intersecting blocks among failed eBCH words is up to $f$, then RS erasure decoding is called to recover those blocks and subsequently the senseword is corrected. We remark that the proposed Phase-I and II decoding is motivated from \cite{Al09}. There is a minor difference such that, reduced-1-bit iterative decoding is run until stalling for Phase-I, as opposed to limiting to the first iteration in \cite{Al09}. In addition, the proposed reduced-1-bit decoding algorithm effectively rules out unfruitful Chien searches.

Phase-III is different from the previous two phases, as each eBCH list decoding may produces multiple candidates. To reduce the number of candidates as well as to reduce search complexity, the evaluation of  $\Delta(x)$, as defined in \eqref{def-Delta}, is limited to the failed intersecting blocks and the parity block.  One trivial solution is to randomly pick a candidate. However, this suffers miscorrection greatly due to high probability of wrong picking.

Herein we present an alternative approach. Upon successfully determining a list of candidates from a failed row (column) constituent word, we perform trial correction on each candidate. Each time check if the crossing column (row) Berlekamp decoding of any previous failed word is successful. We choose the one that results in the most number of crossing column (row) corrections and make both row and column corrections accordingly. We shall discard all if none results in a crossing column (row) correction. Then the next row (column) is carried out in the same manner. This approach effectively takes advantage of crossing validation. Also note this trial-and-error approach is performed at the last phase wherein only a few rows and columns remain to be corrected, so the complexity increment is at most moderate.

Two indicator vectors are exploited to ease implementation, namely, the correction indicator vector and the syndrome update indicator vector. The correction indicator vector tracks correction status. When an eBCH constituent is corrected, its syndromes are reset to zeros, while its corresponding indicator is delayed for a whole iteration to set to 1 (the rationale behind is to exploit cross decoding validation to reduce miscorrection). The proposed iterative decoding skips a constituent eBCH word if its correction indicator is 1. Moreover, the evaluation of $\Delta(x)$ in Phase-III skips blocks that are corrected from earlier iterations, i.e., their correction indicators are 1. The syndrome update vector keeps track of syndrome update. When correction is made by crossing eBCH constituents, its syndromes are updated accordingly and the corresponding indicator is set to 1. The proposed iterative decoding skips a constituent eBCH word if its syndrome update indicator is 0. When decoding of a eBCH constituent is done (Regardless of failure or success), its syndrome update indicator is reset to zero. At the beginning of each phase, syndrome update indicators corresponding to all rows/columns are initialized to 1.

\section{Performance Evaluation}
\label{sec:evaluation}

\begin{figure}
  \includegraphics[width=1\textwidth]{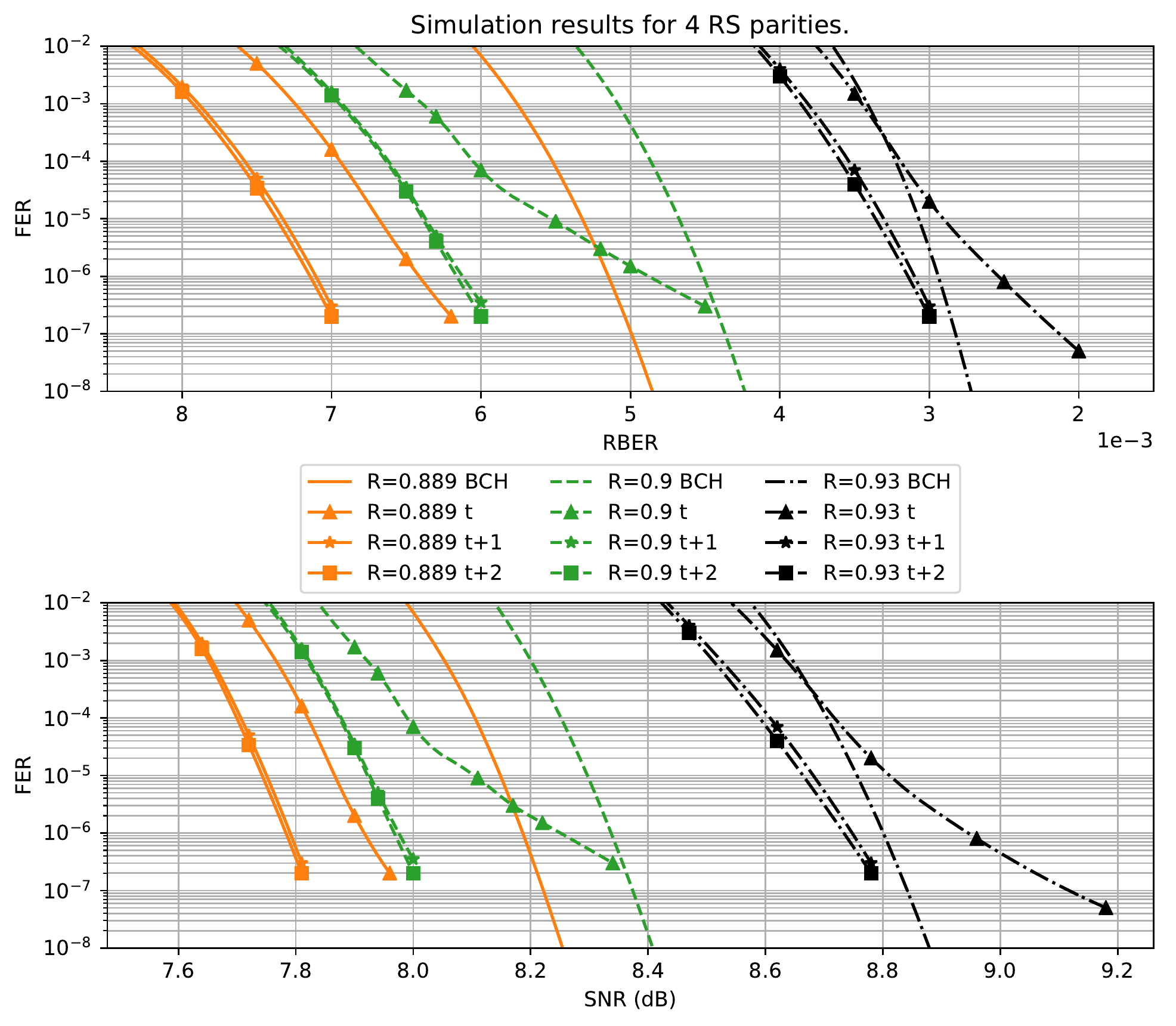}
  \caption{\label{fig:sim} R is the code rate, FER is the frame error rate,
    RBER is the raw bit error rate, and SNR is the signal to noise ratio
    ($2E_b/N_0$). }
\end{figure}

We simulate the proposed codes and decoding algorithms to evaluate their
effectiveness. The size of the user data used for the simulation is 4kB
($K = 32768$). We evaluate code rates of 0.889, 0.9 and 0.93. For each code
rate, we simulate decoding with no list decoding, with up to extra-1-bit
list decoding, and with up to extra-2-bit list decoding. We further compare
with stand-alone BCH codes of the same rates and lengths. The exact numbers
of parity bits are specified in Table ~\ref{Tab:Codelenghts}.

\begin{table}
  \centering
  \caption{Number of parity bits in simulated codes with respect to the fixed data size of
    $K=32768$. Right column shows BCH correcting power
    ($t$).
    \label{Tab:Codelenghts}}
  \begin{tabular}{|l|c|c|c|lll}
    \hline
    Rate & BWP & BCH & $t$ (BCH) \\ \hline
    0.889 & 4082 & 4088 & 258 \\ \hline
    0.9 & 3634 &  3640 & 228 \\ \hline
    0.93 & 2463 & 2472 & 155 \\ \hline
  \end{tabular}
\end{table}

\begin{figure}
  \centering \includegraphics[width=0.7\textwidth]{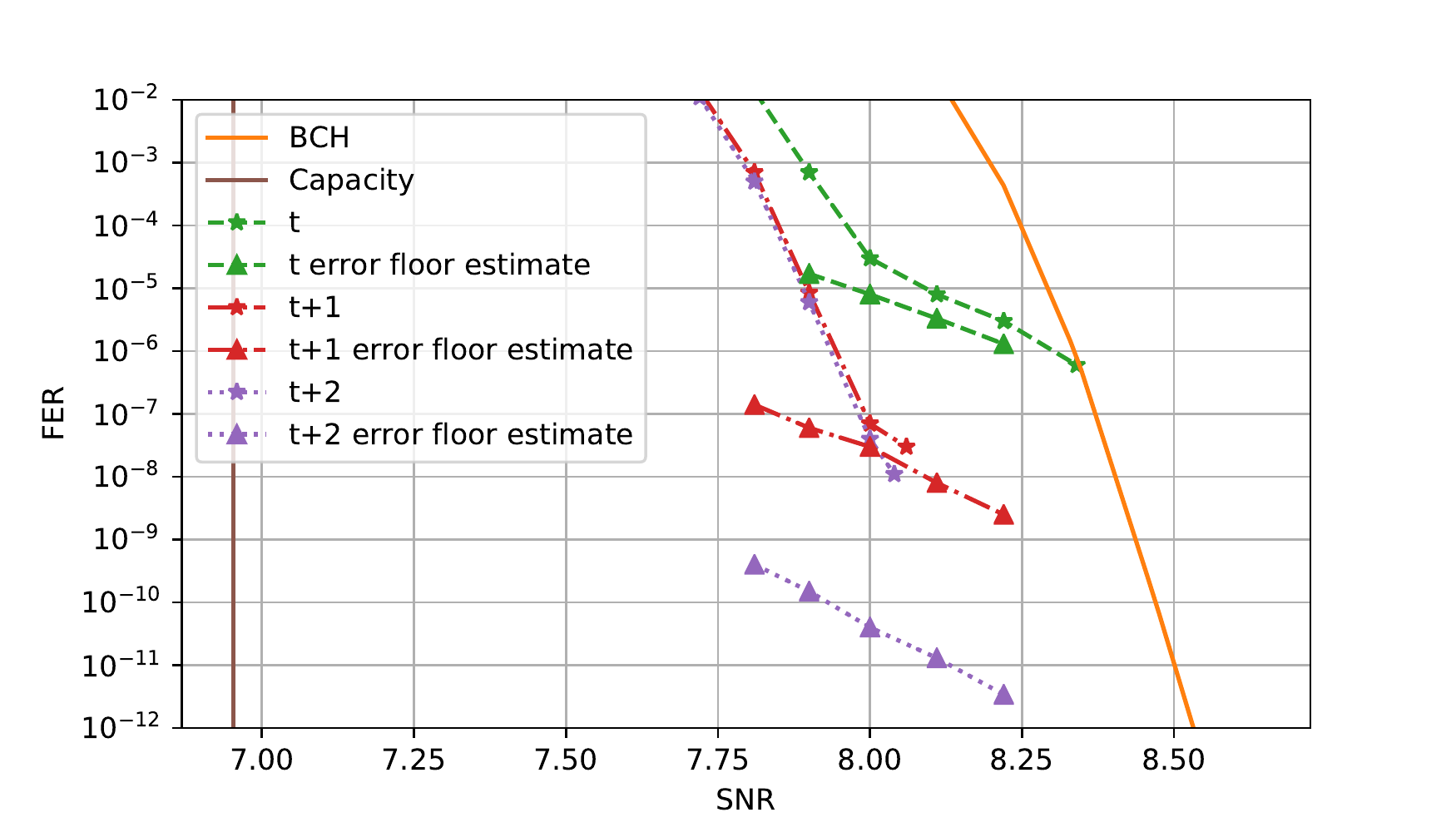}
  \caption{\label{fig:error-floor}Benefit of $+2$ list decoding in error-floor
    region. The error-floor plots are numerical estimates, wherein code rate 0.9, 1
    RS parity, block size $b=20$.}
\end{figure}

\begin{figure}
  \centering \includegraphics[width=0.58\textwidth]{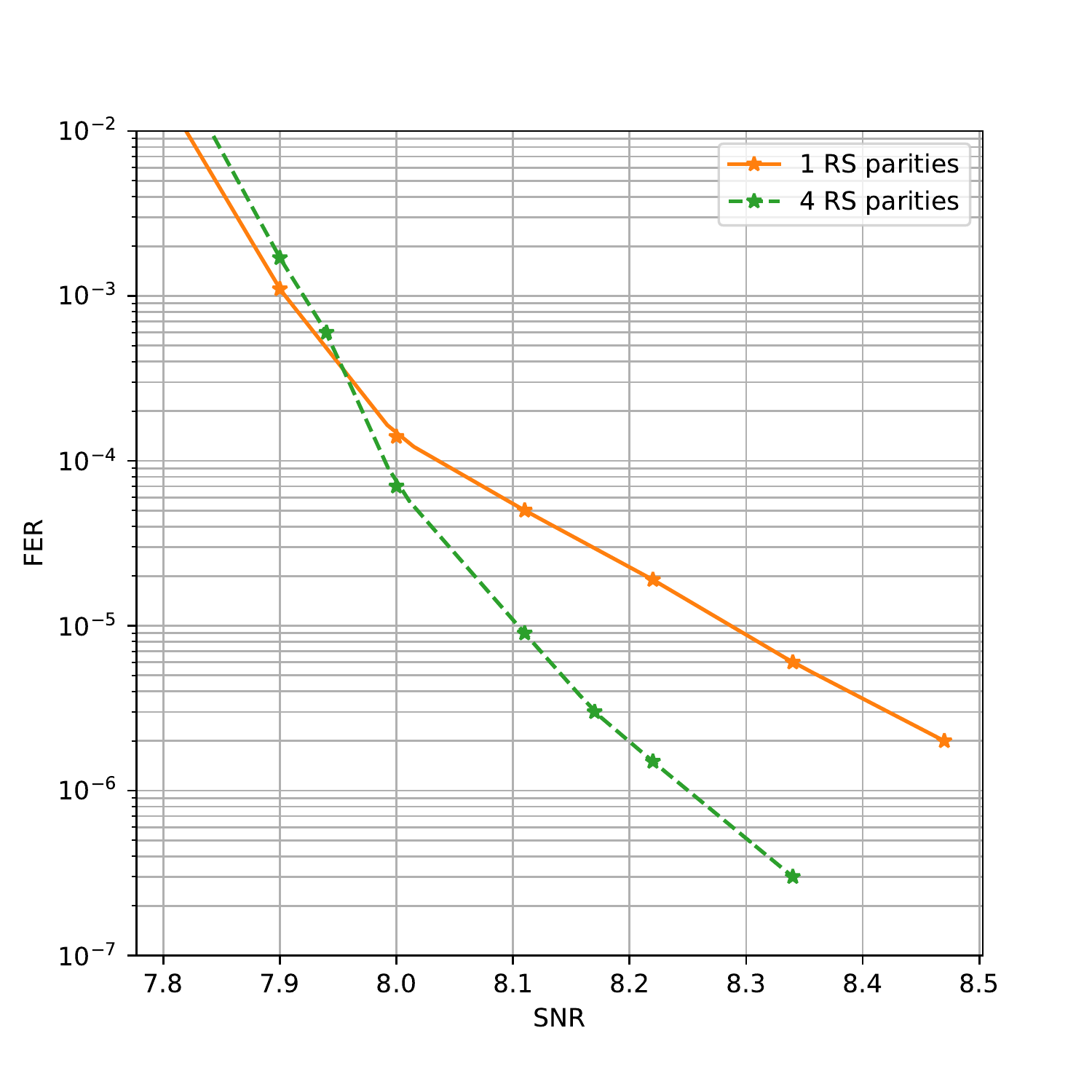}
  \caption{\label{fig:rs-parities}Comparison of 1 vs 4 RS parities with
    unique decoding at code rate 0.9 and block size $b=15$. }
\end{figure}

\begin{figure}
  \centering \includegraphics[width=0.58\textwidth]{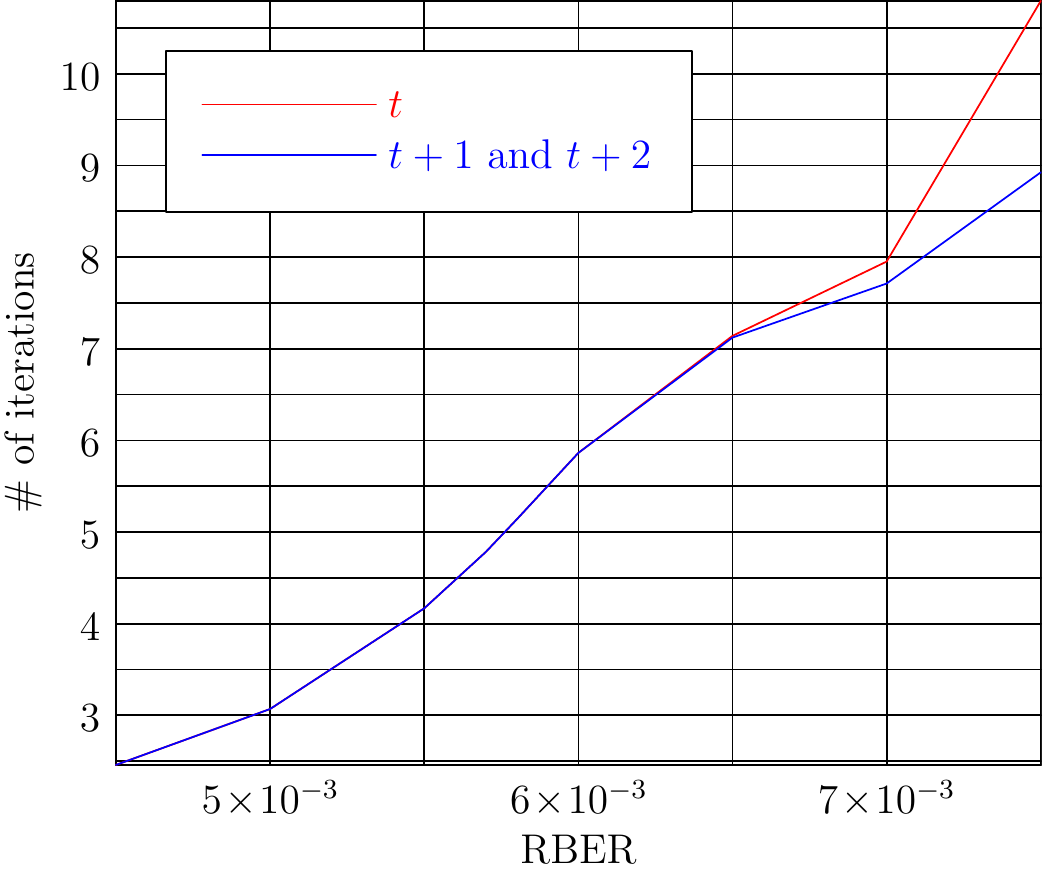}
  \caption{\label{fig:iterations}Mean number of iterations under code rate $0.9$, 1
    RS parity and block size 20. }
\end{figure}

We use an inner RS code with $f=4$ parities to reduce the error floor. The
main additional free parameter is the block size $b$. For each code
rate, we simulate various block sizes to identify one with good
performance. An accurate optimization of the block size would not be
practical, since simulating low frame error rates takes a long time.
Further, we observe that the best performing block size for a given raw-bit-error-rate (RBER) and a given decoding algorithm tends also to perform (near) the best for minor varying RBERs and minor different algorithms. To this end, we settle for a good performing block size for each given code, instead of separately optimizing for each RBER point and each decoding algorithm. The block sizes we find to perform well are $b=20$
for rate 0.889, $b=15$ for rate 0.9 and $b=50$ for rate 0.93. The
simulation results are presented in Figure \ref{fig:sim}. The maximum
number of Turbo iterations is set to 32. Each simulation runs until 100
BWP-BCH frame failures are observed.

The simulation results provide several insights:
\begin{enumerate}
\item List decoding is significantly superior to unique decoding in this
  setting. The error floor of unique decoding is very high, while list
  decoding reduces the error floor below the observable zone.

\item In the lower code rates, BWP-BCH codes provide significant gain over
  stand-alone BCH, with 0.4 dB for rate 0.889. While the gain almost
  entirely disappears in rate 0.93, it is known that other iteratively
  decoded codes, such as LDPC of similar length, also do not outperform BCH
  in this rate (in hard decoding).

\item Incorporating $+2$ list decoding provides a very small gain over that of
  $+1$ list decoding. However, we show next that incorporating $+2$ list decoding is beneficial for
  reducing the error-floor.
\end{enumerate}

To show the benefit of incorporating $+2$ list decoding over that of $+1$ list decoding, we
use a numerical method to estimate the error floor. The method is straightforward, as described in \cite{Cho14}. We evaluate for the code under rate 0.9, with a
single RS parity and block size of 20. Figure~\ref{fig:error-floor} shows
simulation results along with error floor estimations. The error floor
estimation proves to be quite accurate for radius $t$ and $t+1$. The figure
suggests that $+2$ list decoding improves the error floor by 2.5 orders of
magnitudes over $t+1$. On another note, at the benchmark frame-error-rate of $1e-6$, the proposed iterative decoding algorithm is shown to be 1 dB gap from the Shannon capacity, while gaining 0.35 dB over the BCH performance.

Note that the error-floor estimation method from~\cite{Cho14} does not
consider list decoding. To use this method with list decoding ($t+1$ and
$+2$ list decoding), we assume that the list decoding always returns only the
unique (correct) codeword. In other words, we use the extended decoding
radius as if it was the actual decoding radius of the BCH component codes.
The accuracy of the assumption is supported by the relative accuracy of the
estimations for $t$ and $t+1$ decoding.

It remains unclear as to why is the $+2$ list decoding improvement is so
insignificant in the waterfall region. An investigation of the failure
scenarios reveals the reason. In most cases when $+1$ list decoding gets stuck,
the number of non-zero BCH syndromes is rather high. In such case,
miscorrections often happen in the BCH decoding, since the hint from
the opposite dimension is not so helpful. As a result, the BWP
decoding is unable to correct those BCH sensewords.

In contrast, in the error-floor region, the number of non-zero BCH
syndromes is small, resulting in a good hint for the list decoder.
Therefore, $+2$ list decoding resolves most $t+1$ failures. Note that the
reasoning above is in principal also valid for the gain of $t+1$
decoding over $t$ decoding in the waterfall region. However, we do see
a significant gain in $+1$ list decoding in the waterfall region. The
reason for this behavior is that the lists of radius $t+1$ are almost
always much smaller than those of radius $t+2$ for the values of $t$
used in the considered simulations (around $5$).

Finally, note that the code in Figure~\ref{fig:error-floor} contains a
single RS parity, while the codes in Figure~\ref{fig:sim} contain 4
parities. It is instructive to consider the trade-off that governs the
choice for the number of RS parities. Intuitively, we would expect a
code with less RS parities to perform better in the waterfall region,
since the parities are used to strengthen the BCH codes. On the other
hand, a code with more RS parities would perform better in the
error-floor region, since the additional RS parities would protect
better from errors concentrated in a small numbers of blocks. The
choice of the number of RS parities should be made according to the
desired error-floor level. We illustrate the trade-off by the simulation
results in Figure~\ref{fig:rs-parities}, at code rate 0.9 and block
size $b=15$, with unique decoding. Furthermore, the mean number of
iterations is presented in Figure~\ref{fig:iterations}.

\section{Concluding Remarks}

The paper studies BWP-BCH codes with a focus on flash memory applications. Firstly, novel
efficient BCH decoding algorithms are firstly presented, including $-1$ decoding,
$+1$ list decoding, and $+2$ list decoding. Secondly, a unified construction framework
of BWP-BCH codes is presented by leveraging many design freedoms to compromise among design scalability, implementation simplicity and superior performance.
A high-speed scalable encoder is described. Finally, a novel iterative decoding
algorithm for BWP-BCH codes is presented, which utilizes the proposed BCH decoding
algorithms to optimize decoding performance. Simulation results demonstrate superior waterfall performance and significantly lowered error floor. Notably, it achieves 1 dB gap from capacity under the benchmark of 4kB data size, the code rate of 0.9 and the frame-error-rate of 1e$-6$.

There are many problems to be explored. One interesting problem is to build soft information inside the proposed iterative decoding. One would wonder how much further gain may be achieved. Another interesting problem is to extend the proposed decoding algorithm to soft-input soft-output decoding. Furthermore, it is easily observed that concatenating an inner RS code significantly improves the minimum distance bound. However, to our view, it is still too loose to be meaningful in a straightforward manner. Finally, on the hardware perspective, the proposed $+1$ list decoding employs dynamic grouping and rational function evaluation, which are overly complex. A simplified hardware implementation would certainly renders the proposed iterative decoding more practical.

\renewcommand{\baselinestretch}{1.0}

\end{document}